\documentclass[epj]{svjour}
\usepackage{graphics}
\DeclareGraphicsExtensions{.eps,.ps}
\usepackage{layout}
\usepackage[dvips]{epsfig}
\usepackage{amssymb,xspace}
\usepackage{psfrag}
\usepackage{rotate}
\newcommand\sneu{\mbox{$\widetilde{\nu}$}}

\newcommand\stopq{\mbox{$\widetilde{t}$}}

\newcommand\pb{\mbox{\rm pb}}
\newcommand\rp{\mbox{$R_p$}}
\newcommand\rpv{\mbox{$R_p \!\!\!\!\!\! / \;\; $}}

\begin{document}
\title{A Review of Searches for R-parity-violating SUSY}
\author{Christian Schwanenberger 
\thanks{Talk presented at the International Europhysics
  Conference on High Energy Physics EPS 2003, Aachen, July 2003.}
\thanks{{\it Send offprint requests to:}{\tt\ schwanen@mail.desy.de}}
}                     
%
%
\institute{Deutsches Elektronen-Synchrotron (DESY), Notkestr. 85,
D-22607 Hamburg
}
\date{Received: date / Revised version: date}
%
%
\abstract{
Searches for pair
and single production of supersymmetric particles under the assumption that
$R$-parity is violated via a single dominant coupling are presented. A
subset of the most recent results from LEP, Tevatron and HERA is selected. The
data are in 
agreement with the Standard Model expectation. Limits on the production
cross sections and the masses of supersymmetric particles are derived. 
} 
\maketitle
%
%
%
\section{Introduction}
\label{intro}
%
$R$-parity (\rp) \cite{rp} is a discrete multiplicative symmetry. It can be
written as $R_p \equiv (-1)^{3B+L+2S}$.
Here $B$ ($L$) denote the baryon (lepton) number and $S$ the spin of a
particle. For 
Standard Model (SM) particles $\rp=+1$, while $\rp=-1$ for their
supersymmetric partners. 
The most general superpotential with the minimal field content of the
supersymmetric SM contains the trilinear \rp-violating terms\footnote{Bilinear
  terms are 
  not considered in this report.}: 
\begin{equation}
\label{wrpv}
{W_{\rpv} = {\lambda_{ijk}} L_i L_j
   \bar{E}_k + {\lambda'_{ijk}} L_i Q_j \bar{D}_k + {\lambda''_{ijk}}
 \bar{U}_i \bar{D}_j \bar{D}_k } \,\, ,
\end{equation}
where $L_{i,j}$ ($Q_j$) are the lepton (quark) doublet superfields,
$\bar{D}_{j,k}$, 
$\bar{U}_i$ ($\bar{E}_k$) are the down-like and up-like quark (lepton) singlet
superfields, $\lambda$, $\lambda'$ ($\lambda''$) are Yukawa couplings which
violate $L$ ($B$) conservation\footnote{Fast proton decay is suppressed if
  $\Delta L \neq 0$ and $\Delta B 
  \neq 0$ operators are not simultaneously present.} and
$i,j,k=1,2,3$ are the generation indices \cite{yuk}.\footnote{Here 45 \rpv\
couplings are introduced. If spontaneous \rpv\ 
is considered, there are three additional terms.} 
Hierarchies in \rpv\
couplings are supposed as for Yukawa couplings generating fermion
masses. Thus, here one assumes that a single coupling dominates.

\rpv\ opens a new scenario in supersymmetric searches complementary to the
\rp-conserving ones.
The conservation of $R_p$ implies that sparticles
can only be produced in pairs. Their decay ends in final states containing
the lightest 
supersymmetric particle (LSP) which is stable.
The phenomenological consequences of \rpv\ are that
single sparticle production is possible and that the LSP 
decays into SM fermions. The latter can lead to final states with a large
number of leptons or jets.

In the analyses presented here both single sparticle production via a \rpv\
coupling and pair production of sparticles via a \rp-conserving coupling are
investigated. The study of pair-produced sparticles allows one to
constrain
sparticle masses, independently of whether $R_p$ is conserved. 
In the analyses presented here two types of sparticle decays are
considered. First, the {\em direct decay} into two
SM fermions via a \rpv\ vertex and, second, the {\em
  indirect decay} via \rp-conserving interactions in
cascades down to the LSP, which then decays via a \rpv\ vertex.
%
%
%
\section{LEP searches}
\label{sec:lep}
\begin{figure}
\hspace{0.5cm}
\resizebox{0.4\textwidth}{!}{%
 \psfrag{(a)}[][][1.5][0]{ }
{\epsfig{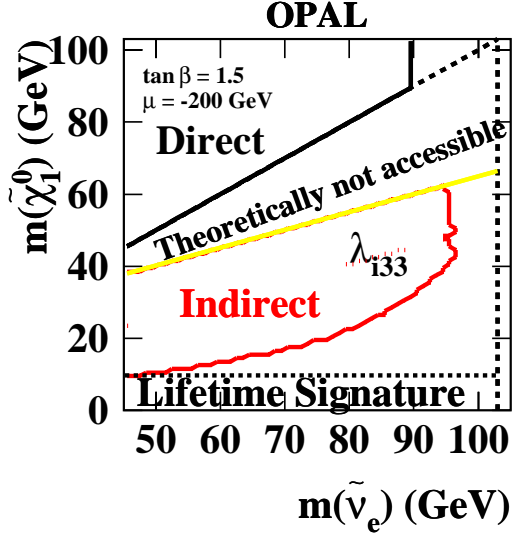}}
}
\vspace{7.2cm}       
\caption{CMSSM exclusion region for $\sneu_e \sneu_e$ production 
in the ($m_{\sneu_e}$, $m_{\widetilde{\chi}_1^0}$) plane 
at $95\%$ confidence level
(CL) for a $\lambda$ coupling}
\label{fig:1}       
\end{figure}

%
%
In $e^+ e^-$ collisions sparticle pair production has been investigated
at center 
of mass energies up to 209~GeV. No deviation
from the SM was found. 
Limits are derived on 
cross-sections and couplings $\lambda$, $\lambda'$, $\lambda''$. The limits
on sparticle masses are as stringent as in the
\rp-conserving analyses. Furthermore, exclusion plots in the 
Minimal Supersymmetric Standard Model (MSSM) are produced. 
Details can be found in \cite{lep}. 
As an example, exclusion limits in a Constrained MSSM (CMSSM) for 
$\sneu_e$ pair production for $\sneu_e$ decaying via 
a $\lambda$ coupling ({\em direct}) into 4 leptons ($e^+ e^- \to \sneu
\bar{\sneu} \to  \ell^- \ell^+ \, \ell^+ \ell^-$) or via a gauge coupling
({\em indirect}) producing 4 leptons and missing energy ($e^+ e^- \to \sneu
\bar{\sneu} \to \nu {\widetilde{\chi}_1^0} \bar{\nu} {\widetilde{\chi}_1^0}
\to \nu (\stackrel{(-)}{\nu} \ell^+ \ell^-) \ \bar{\nu} (\stackrel{(-)}{\nu}
\ell^+ \ell^-)$ ) are shown in
Fig.~\ref{fig:1}. This analysis was performed by the OPAL experiment using
a total 
integrated luminosity of $610\ {\rm pb}^{-1}$.

%
%
%
%
\section{Tevatron searches}
\label{sec:tevatron}
%

\begin{figure}
\vspace{-0.3cm}
\hspace{0.5cm}
\resizebox{0.4\textwidth}{!}{%
{\epsfig{file=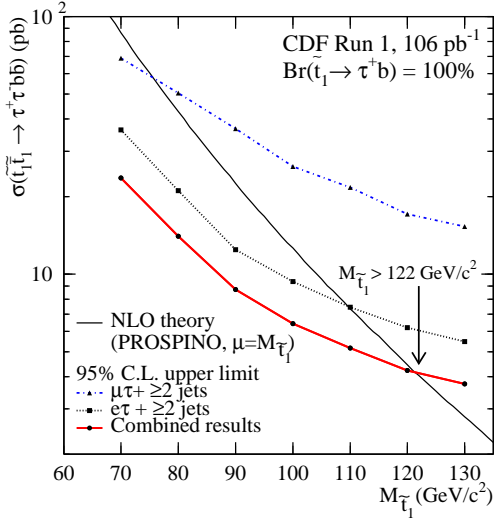,bbllx=0,bblly=540,bburx=540,bbury=0}}
}
\vspace{7cm}       
\caption{Upper limits at $95\%$ CL on cross section for $\stopq_1
  \bar{\stopq}_1$ production compared to the NLO calculations}
\label{fig:2}       
\end{figure}

A search for pair production of
stop quarks ($\stopq_1$) using $106\ \pb^{-1}$ of $p\bar{p}$ collisions at
$\sqrt{s}=1.8$~TeV has been performed by CDF \cite{Acosta:2003ze} (Tevatron Run
I). In the 
investigated mode each 
$\stopq_1$ decays into a $\tau$ lepton and a $b$ quark. The search demands
events with two $\tau$'s, one decaying leptonically ($e$ or $\mu$) and one
decaying hadronically, and two jets. No candidate event passes the final
selection criteria.
Fig.~\ref{fig:2} shows upper limits on the cross section
for a branching ratio of $100\%$ for the $e$, $\mu$ and combined channels,
along with 
the NLO prediction of the production cross sections. A lower limit of
$\approx 122$~GeV for the $\stopq_1$ 
mass has been derived. Since the
analysis does not 
distinguish the quark flavors in jet reconstruction, these results are
equally valid for any $\lambda'_{33k}$ ($k=1,2,3$) coupling. 
%
%
%
\section{HERA searches}
\label{sec:hera}
%
%
The $ep$ collider HERA is well suited to search for resonant squark
production via \rpv\ 
interactions. 
With an 
initial $e^+$ beam the sensitivity is highest to couplings $\lambda'_{1j1}$
($j=1,2,3$), where mainly 
$\widetilde{u}_L^j$ 
squarks are produced. In
contrast, with an initial $e^-$ beam HERA is most sensitive to couplings
$\lambda'_{11k}$ ($k=1,2,3$) and can mainly produce $\widetilde{d}^k_R$
squarks.

The resonant production of single $\widetilde{u}_L^j$ and
$\widetilde{d}_R^k$ squarks has been investigated by H1 using data taken at a
center of 
mass energy of 
$\sqrt{s}=319$~GeV corresponding to an integrated luminosity of
$64\ \pb^{-1}$ for $e^+p$ collisions and $13.5\ \pb^{-1}$ for $e^-p$
collisions \cite{squarks}.
No deviation from the SM prediction has been observed.
Upper bounds on the production cross section are
derived by combining all channels from {\em direct} and {\em
  indirect} squark decays. In SUSY models inspired by the Minimal
Supersymmetric Standard Model (MSSM), the channels considered in this
analysis cover $\approx 100\%$ of all possible squark decay modes. These
bounds are 
translated into constraints on the parameters of SUSY models. 
%
%
\begin{figure}
\hspace{1cm}
\resizebox{0.35\textwidth}{!}{%
  {\epsfig{file=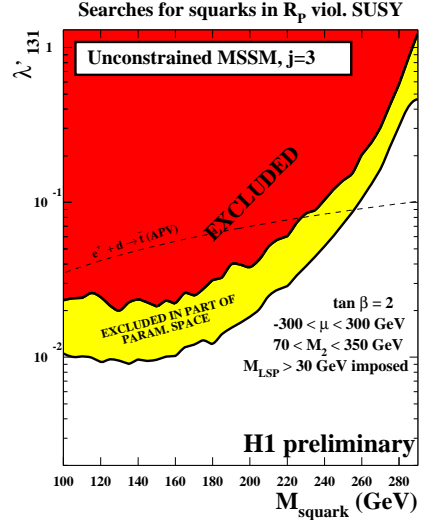,bbllx=40bp,bblly=690bp,bburx=530bp,bbury=130bp,clip=}}
}
\vspace{7cm}       
\caption{Upper limits at $95\%$ CL for the coupling $\lambda'_{131}$ as a function of
  the squark mass, in the ``phenomenological'' MSSM}
\label{fig:3}       
\end{figure}
%
%
%
\begin{figure}
\vspace{1.5cm}
\resizebox{0.45\textwidth}{!}{%
{\epsfig{file=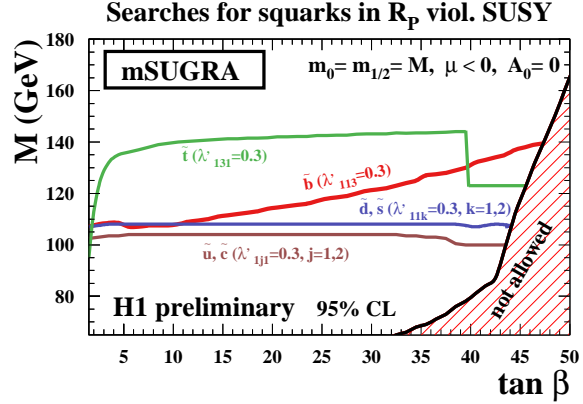,bbllx=39,bblly=491,bburx=503,bbury=239}}
}
\vspace{4cm}       
\caption{Constraints at $95\%$ CL obtained in the mSUGRA model, assuming $\lambda'_{1jk}
  =0.3$ ($j,k=1,2,3$). The areas below the curves are excluded}
\label{fig:4}       
\end{figure}
%

Fig.~\ref{fig:3} shows upper limits on the Yukawa coupling $\lambda'_{131}$
as a function of the $\stopq_L$ mass. These are obtained in a
``phenomenological'' MSSM, where the gaugino masses are related to each
other while the sfermion masses are free. 
A scan of the parameter space is
performed, which shows that the obtained limits do not depend strongly on
the model parameters. The limits
obtained on the coupling $\lambda'_{131}$ (and on $\lambda'_{121}$ --- 
not shown here) extend beyond the indirect bounds from low energy
experiments.
For a coupling of electromagnetic strength
($\lambda'=0.3$) stop masses up to $\approx 270$~GeV are excluded. E.g. the
future sensitivity of the Tevatron Run II experiments on light stop quarks
might 
be around 200-250 GeV, depending on the main decay modes of the stop. 
Thus, a
reasonably large coupling $\lambda'_{131}$ would provide an
interesting discovery potential for the stop at HERA II with the much larger
integrated luminosity expected within the next few years. 

Constraints obtained in the framework of the minimal Supergravity
(mSUGRA) model have also been derived. 
Here, a common mass $m_0$
($m_{1/2}$) is assumed for the scalars (gauginos) at the Grand Unification
scale. 
Lower limits for $m_0=m_{1/2}=M$ ($\lambda'=0.3$) are presented in
Fig.~\ref{fig:4} as a function of
$\tan\beta$, the ratio of the vacuum expectation values of the two neutral 
scalar Higgs fields. The single production of all 6 squark flavors is
considered.  
%
%

In another H1 analysis, neutralino ($\widetilde{\chi}_1^0$) production via 
\rpv\ $t$-channel selectron exchange ($e^+ p \to
\widetilde{\chi}_1^0 + {\rm 1\, jet}$)
has been
studied in Gauge Mediated Supersymmetry Breaking (GMSB) scenarios where the
slepton 
masses are usually much lower than the squark 
masses \cite{gravitino}. The gravitino is the LSP, while the
neutralino is 
assumed to be the 
next-to-lightest supersymmetric particle (NLSP).
The prompt
decay of the neutralino into a photon and a gravitino is analyzed, which
leads to prominent event signatures with a photon and large missing
transverse momentum. 
No significant
deviation from the SM has been found using
$64\ \pb^{-1}$ of $e^+p$ collisions at $\sqrt{s}=319$~GeV.

\begin{figure}
\hspace{0.75cm}
\resizebox{0.37\textwidth}{!}{%
   \psfrag{x}[][][1.5][0]{$\bf\widetilde{\chi}^0_1$}
{\epsfig{file=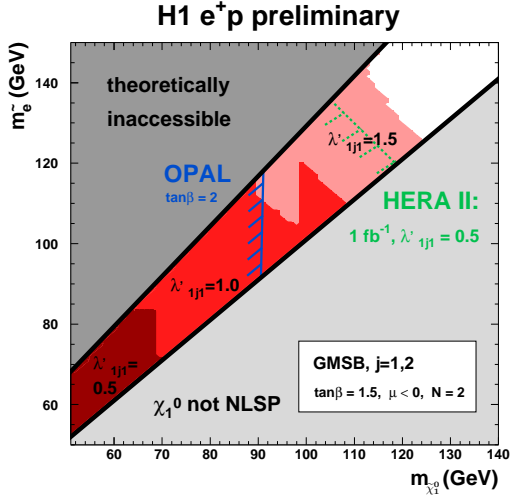,bbllx=25,bblly=660,bburx=575,bbury=130}}
}
\vspace{6.1cm}       
\caption{Excluded regions at $95\%$ CL in the GMSB model dependent on selectron and neutralino mass assuming $\lambda'_{1j1} =$ 0.5 (dark red), 1.0 (red), 1.5 (light
red) ($j=1,2$) 
}
\label{fig:5}       
\end{figure}
%
%
In Fig.~\ref{fig:5}, excluded regions in the
($m_{\tilde{e}}$,$m_{\widetilde{\chi}^0_1}$) plane for 
different 
values of the \rpv\ coupling are 
presented.
For selectron masses very close to the neutralino mass and for
$\lambda'_{1j1} = 1.0$, neutralino masses up to $\approx 108$ GeV can be excluded.
For moderate values of $\lambda'_{1j1}$, the region excluded by the OPAL
\cite{Abbiendi:2000bd} analysis of $e^+ 
e^- \rightarrow \widetilde{\chi}^0_1 \widetilde{\chi}^0_1 \rightarrow \gamma \tilde{G} \gamma
\tilde{G}$ in a \rp-conserving SUSY scenario can be extended.
%
\begin{figure}
\hspace{0.75cm}
\resizebox{0.37\textwidth}{!}{%
   \psfrag{x}[][][1.5][0]{$\bf\widetilde{\chi}^0_1$}
  {\epsfig{file=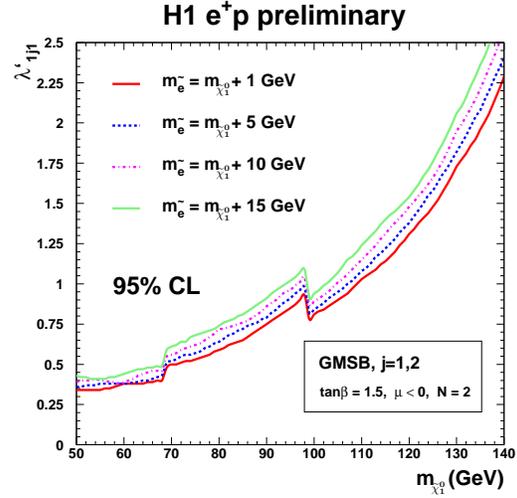,bbllx=25,bblly=660,bburx=575,bbury=130}} 
}
\vspace{6.1cm}       
\caption{Upper limits at $95\%$ CL for the coupling
    $\lambda'_{1j1}$ ($j=1,2$) as a function
    of the neutralino mass for various differences
    between the selectron and neutralino masses
}
\label{fig:6}       
\end{figure}
%
In Fig.~\ref{fig:6}, upper limits on
$\lambda'_{1j1}$ ($j=1,2$) are given as a function of $m_{\widetilde{\chi}^0_1}$ for various
assumptions for the difference between selectron and neutralino
mass. These are the first limits derived for
$\lambda'_{121}$ which are independent of squark masses.
%
\section{Conclusions and prospects}
\label{sec:concl}
%
The violation of R-parity has inspired new interesting scenarios for SUSY
searches. Many different channels for single or pair production of SUSY
particles in 
a \rpv\ scenario have been investigated at LEP, Tevatron and
HERA. However, no
deviation from the SM has been observed. The derived limits from \rpv\
searches are comparable with the $R_p$-conserving limits. Thus, searches
for \rpv\ deliver important contributions to constraints on
SUSY models. Further interesting results can be expected at Tevatron
Run~II, HERA~II and future colliders.

\vspace{-0.09cm}
%
%
%

\end{document}